%Paper: cond-mat/9408018
%From: SATHYA <GURUSWAMY@urhep.pas.rochester.edu>
%Date: Thu, 04 Aug 1994 12:53:22 -0500 (EST)

%%%%%%%%%%%%%%%%%%%%%%%%%%%BEGIN HERE%%%%%%%%%%%%%%%%%%%%%%%%%%%%%%%%%%
\headline={\ifnum\pageno=1\firstheadline\else
\ifodd\pageno\rightheadline \else\leftheadline\fi\fi}
\def\firstheadline{\hfil}
\def\rightheadline{\hfil}
\def\leftheadline{\hfil}
        \footline={\ifnum\pageno=1\firstfootline\else\otherfootline\fi}
\def\firstfootline{\rm\hss\folio\hss}
\def\otherfootline{\hfil}

\font\tenrm=cmr10

\font\elevenbf=cmbx10 scaled\magstep 1
\font\elevenrm=cmr10 scaled\magstep 1
\font\elevenit=cmti10 scaled\magstep 1

\font\ninerm=cmr9

\nopagenumbers
\hsize=6.0truein
\vsize=8.5truein
\parindent=1.5pc
\baselineskip=10pt
%PAPER BEGINS HERE

%Reference and equation counters

\def\ifundefined#1{\expandafter\ifx\csname
#1\endcsname\relax}

\newcount\eqnumber \eqnumber=0
\def\beq{ \global\advance\eqnumber by 1 $$ }
\def\eeq{ \eqno(\the\eqnumber)$$ }
\def\label#1{\ifundefined{#1}
\expandafter\xdef\csname #1\endcsname{\the\eqnumber}
\else\message{label #1 already in use}\fi}
\def\(#1){(\csname #1\endcsname)}
\def\puteqno{\global\advance \eqnumber by 1 (\the\eqnumber)}

\def\label#1{
\ifundefined{#1}
      %define new macro whose expansion is eqnumber.
\expandafter\edef\csname #1\endcsname{\the\eqnumber}
\else\message{label #1 already in use}
\fi{}}
\def\(#1){(\csname #1\endcsname)}
\def\eqn#1{(\csname #1\endcsname)}

\newcount\refno \refno=0
\def\[#1]{\ifundefined{#1}\advance\refno by 1
\expandafter\xdef\csname #1\endcsname{\the\refno}
\fi[\csname #1\endcsname]}
\def\refis[#1]{\item{\csname #1\endcsname.}}

\def\sqr#1#2{{\vcenter{\vbox{\hrule height.#2pt
        \hbox{\vrule width.#2pt height#1pt \kern#1pt
          \vrule width.#2pt}
        \hrule height.#2pt}}}}
\def\square{\mathchoice\sqr68\sqr68\sqr{4.2}6\sqr{2.10}6}

\def\n{\noindent}

%PREPARATION TO START

\baselineskip=18pt
%\magnification=1200

%MATH MODE SYMBOLS
\def\d#1/d#2{ {\partial #1\over\partial #2} }

%ASSORTED COUNTERS
\newcount\sectno

\newcount\subsectno

\def\subsect{\global\advance\subsectno by1 \the\sectno.\the\subsectno }
 \def\sect{\subsectno=0 \global\advance\sectno by1 \the\sectno }

%ABBREVIATONS

\def\half{{1\over 2}}

\def\linebreak{\hfil\break}

%EQUATION ENVIRONMENTS

\newcount\eqnumber
\def\beq{ \global\advance\eqnumber by 1 $$ }
\def\eeq{ \eqno(\the\eqnumber)$$ }
\def\puteqno{
\global\advance \eqnumber by 1 (\the\eqnumber)}

%REFERENCE AND EQUATION  COUNTERS

\def\ifundefined#1{\expandafter\ifx\csname
#1\endcsname\relax}
          %checks to see if the arguments is the name of a
          %macro already
          %deined.

\newcount\refno \refno=0  %counts references in order of
                         %apperance in text
\def\[#1]{
\ifundefined{#1}
      %define new macro whose expansion is refnumber.
\advance\refno by 1
\expandafter\edef\csname #1\endcsname{\the\refno}
\fi[\csname #1\endcsname]}
\def\refis#1{\noindent\csname #1\endcsname. }

\def\label#1{
\ifundefined{#1}
      %define new macro whose expansion is eqnumber.
\expandafter\edef\csname #1\endcsname{\the\eqnumber}
\else\message{label #1 already in use}
\fi{}}
\def\(#1){(\csname #1\endcsname)}
\def\eqn#1{(\csname #1\endcsname)}

%PREPARATION TO START
\baselineskip=15pt
%\parskip=10pt

%\magnification=1200

%%%%%%%%%%%%%%%%%%%%%%%%%%%%%%%%%%%%%%%%%%%%%%%%%%%%%%%%%%%%%%%%%%%%%%%%%%
						\hfill UR-1368/ER-40685-818

%==========================================================================
%\centerline{\elevenbf WORLD SCIENTIFIC PUBLISHING COMPANY}
\vglue 7pt
\centerline{\elevenbf A THREE-DIMENSIONAL CONFORMAL FIELD THEORY}
%\vglue 7pt
%\centerline{\elevenbf MANUSCRIPT USING COMPUTER SOFTWARE}
%==========================================================================
\vglue 1.0cm
\centerline{\elevenrm
S. GURUSWAMY$^\dagger$\footnote{*}{\ninerm\baselineskip=11pt
Talk presented by S.G. at the 16$^{\rm th}$ Annual
Montreal-Syracuse-Rochester-Toronto (MRST) Meeting:
``What Next?  Exploring the Future of High-Energy Physics'', held at
McGill University, Montreal, Canada, 11--13 May 1994.
To appear in Proceedings published by World Scientific.},
S.G. RAJEEV$^\dagger$ and P. VITALE$^{\dagger\dagger}$}
\baselineskip=13pt
\vglue 0.3cm
\centerline{\elevenit $^\dagger$ Department of Physics, University of
Rochester}
\baselineskip=12pt
\centerline{\elevenit Rochester, NY 14627, USA}
%    IF NO SECOND AUTHOR, COMMENT OUT FOLLOWING LINES ----------
\vglue 0.1cm
%   \centerline{\elevenrm and}
 %  \vglue 0.3cm
%   \centerline{\elevenrm SECOND AUTHOR'S NAME}
 \centerline{\elevenit
{$^{\dagger\dagger}$ Dipartimento di Scienze Fisiche, Universit\`a di Napoli}}
\baselineskip=12pt
\centerline
{\elevenit {and I.N.F.N. Sez.\ di Napoli,}}
\baselineskip=12pt
\centerline
{\elevenit {Mostra d'Oltremare Pad.~19, 80125 Napoli ITALY}}

%==========================================================================
\vglue 0.8cm
\centerline{\tenrm ABSTRACT}
\vglue 0.3cm
  {\rightskip=3pc
 \leftskip=3pc
 \tenrm\baselineskip=12pt
 \noindent
This talk is based on a recent paper$^{1}$ of ours.
In an attempt to understand
three-dimensional conformal field theories, we study in detail one
such example --
the large $N$ limit of the $O(N)$
non-linear sigma model at its non-trivial fixed point -- in the zeta function
regularization.
We study this on various three-dimensional manifolds
of constant curvature of the kind $\Sigma \times R$
($\Sigma=S^1 \times S^1, S^2,
H^2$). This describes a
quantum phase transition at zero temperature. We illustrate
that the factor that determines whether $m=0$ or not at the
critical point in the different cases is not the `size' of $\Sigma$ or its
Riemannian curvature, but the
 conformal class of the metric.

\vglue 0.8cm }
%==========================================================================
%%%%%%%%%%%%%%%%%%%%%%%%%%%%%%%BEGIN HERE%%%%%%%%%%%%%%%%%%%%%%%%%%%%%%%%%%%%%%
\line{\elevenbf 1. Introduction \hfil}
\vglue 0.2cm
\baselineskip=14pt
\elevenrm

We begin in this paper a study of three-dimensional conformal field theories.
In three dimensions, it is necessary to study field theories in curved
space to understand conformal invariance (see appendix B in Ref.~1).
The particular example we study is the $O(N)$ non-linear sigma model$^{2}$,
in the
 large $N$ limit. This is an interacting theory with a non-trivial UV fixed
point.
We compute the critical properties of this
theory on manifolds $(M,g)$ of the kind $M=\Sigma \times R$, $\Sigma$
being a two-dimensional manifold of constant curvature.
The results can be
summarized as follows:
on $S^1 \times S^1 \times R$ (zero curvature),
the physical mass of the field, $m$, is non-zero at criticality.
In this case, $m$ is the inverse correlation length and therefore
the correlation length is
finite at the critical point. This is due to the finite size of the
manifold in some directions. (Another well
studied$^{3}$ case is $R^2 \times S^1$. In Ref.~1, we rederive
the results in Ref.~3 in the zeta function regularization.)
More generally, whenever the manifold is
not conformally equivalent to $R^3$, we should expect the value of $m$
at criticality to be non-zero and our computations confirm this.
On $S^2 \times R$ (constant positive curvature),
which is conformally equivalent to $R^3-\{0\}$,
the mass $m$ goes to zero at criticality.
$m=0$ does not however mean an infinite correlation length on $S^2 \times R$;
the correlation length at criticality is in fact finite.
This is in contrast with the case $H^2\times R$ (constant
negative curvature) where,
$m\neq 0$ at criticality,
but the correlation length is infinite. Also,
 negative curvature induces the $O(N)$ symmetry to be spontaneously broken at
the critical point.
Thus we see that what matters,
for $m$ to be zero or otherwise, is the
conformal class of the metric and not the `size' of the system.
(Limitations of space force us to merely state the results in several places.
 The
interested reader is referred to the longer version of this paper$^{1}$
for a general discussion on conformal invariance,
for technical details and for a complete set of references.)
\vglue 0.6cm
\line{\elevenbf 2. Large $N$ limit of the $O(N)$ non-linear sigma model
in three dimensions \hfil}
\vglue 0.3cm
The regularized euclidean
partition function of the $O(N)$ non-linear sigma model in three
dimensions, in the
 presence of a background metric, $g_{\mu \nu}(x)$,
 can be written as,
\beq {\cal Z}[g, \Lambda, {\lambda(\Lambda)}]=\int {\cal D}_{\Lambda}[\phi]
{\cal D}_{\Lambda}[\sigma]
{\rm exp}\bigl\{{-\int {d^3x
{\surd g}[{1\over 2}{\phi^i}(-\square_g+{\sigma}) {\phi_i}
-{{\sigma}\Lambda \over 2{\lambda(\Lambda)}}]}}\bigr\}\eeq
where $i=1,2,\cdots,N$. $\lambda$ is a coupling constant
and $\Lambda$ is the ultraviolet cut-off introduced to regulate the theory.
(${\cal D}_{\Lambda}[\phi]
=\prod\limits_{|k|<{\Lambda}} d\phi (k)$
and similarly ${\cal D}_{\Lambda}[\sigma]$.)
The constraint on the $\phi$ fields, $\phi^i(x) \phi_i(x)=1$,
has been implemented by a
Lagrange multiplier, in the form of an auxiliary field $\sigma(x)$.
(The canonical dimensions of $\sigma(x)$ in mass units is,
$[\sigma(x)]=2$.)
The part of the action quadratic in $\phi^i$ is
conformally invariant$^{4}$ under the conformal transformation
of the metric,
$g^{\mu \nu}(x)\rightarrow \Omega^{2}(x) g^{\mu \nu}(x)$ with
$\phi(x)\rightarrow \Omega^{1-{d\over 2}}(x) \phi(x)$ and
$\sigma(x) \rightarrow\Omega^{-2}(x) \sigma(x)$.
$-\square_g$ is the ``conformal laplacian'',
$-\square_g=
{-{1\over{\surd g}}\partial_{\mu}({\surd g(x)}g^{\mu\nu}(x)\partial_{\nu})}
+\xi {\rm R}$ where ${\rm R}$ denotes the
Ricci scalar and
$\xi={{d-2}\over {4 (d-1)}}$ and is equal to $1/8$ for dimensions $d=3$.
Although the classical action
is not conformally invariant,
we will see that there
is a non-trivial fixed point for the quantum theory at which ${\cal Z}$
is conformally
invariant.

We will now study this problem in the large $N$ limit, as it is not
possible to solve this theory exactly.
In the large $N$ limit ({\elevenit i.e.}, $N\rightarrow \infty$ with
$N {\lambda(\Lambda)}$ fixed), the generating functional can be calculated
using the saddle point approximation. To do this,
redefine
$(N-1) \lambda(\Lambda)$ in the action as $\lambda(\Lambda)$ which we keep
fixed as $N\rightarrow \infty$. Also, rescale the field $\phi_N$ to
${\sqrt {N-1}}\phi_N$. On spaces of constant curvature,
we can integrate out the first $N-1$ components of the $\phi$ field and
rewrite $\cal Z$ as,
\beq\eqalign{& {\cal Z}[g,\Lambda,\lambda(\Lambda)]\cr&=
\int{\cal D}_{\Lambda}[\phi_N]
{\cal D}_{\Lambda}[\sigma]
{\rm exp}\Bigl\{{{-{(N-1)\over 2}\bigl[{\rm TrLog}_{\Lambda} (-\square_g+
{\sigma})
+\int d^3x {\surd g}
[{\phi_N}(-\square_g+{\sigma}) {\phi_N}
-{\Lambda \over {\lambda(\Lambda)}}{\sigma(x)}]\bigr]}}\Bigr\}\cr}\eeq

In the examples we consider in our work, the manifolds are of constant
curvature and we can therefore look for the uniform saddle point,
$\sigma(x)=m^2$ and $\phi_{N}(x)=b$.
(The physical interpretation for $m$ is that of the physical mass of
the $\phi$ field and
for $b$ is that
of the vacuum expectation value of the $\phi$ field or spontaneous
magnetization.)
The saddle points are the solutions to the following
``gap equations'', which are obtained by extremizing
the action with respect to
$\phi_{N}(x)$ keeping $\sigma(x)$ fixed and vice--versa:
\beq (-\square_g+m^2) b=0\eeq \label{gap1}
\beq
b^{2}= {\Lambda \over {\lambda(\Lambda)}}
- G_{\Lambda}(x, x; m^2,g),\eeq \label{gap2}

\n where $G_{\Lambda}(x, x; m^2,g)
=\langle x|(-\square_g+ m^2)^{-1}|x \rangle_
{\Lambda}$. $G_{\Lambda}(x^{\prime},x;m^2,g)$ is the
two point correlation function of the $\phi$ fields.

We are interested in studying the critical behaviour of our theory. This is
the point at which the theory becomes conformally invariant and all the
divergences in the theory vanish. The question then is:
Do the divergences that arise
in the Green's function, $G_{\Lambda}(x, x; m^2,g)$, depend on the
metric $g^{\mu \nu}$?
%We establish here that the only divergences that arise in
%the Green's function in curved space are those which
%arise on $R^3$.
\beq {G_{\Lambda}(x, x; \sigma,g)
=\int\limits_{1\over \Lambda^2}^{\infty} dt
{\langle x|e^{-(-\square_g+\sigma)t}|x \rangle}}
=\int\limits_{1\over \Lambda^2}^{\infty} dt \quad h(t;x,x)\eeq
where $h(t;x,y)$ is the heat kernel of the operator
$-\square_g+\sigma$.
The divergence in the Green's function comes from short distance (UV)
and hence
from the small $t$ region. To isolate the divergent part of the Green's
function, from the finite part, we use the
asymptotic expansion of the heat kernel$^{1}$, which,
in three dimensions, is given by,
\beq h(t;x,y)={e^{-{d^2(x,y) \over 4 t}} \over (4 \pi t)^{3 \over 2}}
\sum_{n=0}^\infty a_n(x,y) t^n\eeq where $d(x,y)$ is the Riemannian distance
 between points $x$ and $y$ on the manifold $M$.
The leading term, with $a_0(x)=1$, is the only term that produces a divergence
in the Green's function:
$$G_{\Lambda}(x, x; \sigma,g)={\Lambda \over 4(\pi)^{3\over 2}} +
{\rm finite ~part}.$$
We see that the divergent part of the Green's function is independent
of the metric $g^{\mu \nu}(x)$. Thus the critical value of the coupling
constant, $\lambda_c(\Lambda)$,
at which the theory becomes finite, is independent of the background metric.
\vglue 0.6cm
\line{\elevenbf 3. Zeta function regularization \hfil}
\vglue 0.3cm

We will be using the zeta function
regularization in this work as we find it to be more
tractable when we are in curved space.
Since the critical value of the coupling constant
at which the theory becomes finite is independent of the metric, we
find this critical coupling on $R^3$.
The Green's function, $G(x,x;m^2,g)$, is regularized as follows:
\beq G(x,x;m^2,g)=\lim_{s \rightarrow 1}
\langle x|(-\square_g+m^2)^{-s}|x \rangle
=\lim_{s \rightarrow 1} \zeta_g(s).\eeq
where
$\zeta_g(s)
=\sum_{\lambda_n \not=0} |\lambda_n|^{-s}$ with $\lambda_n$ as
the eigenvalues of $(-\square_g+m^2)$ and the sum includes multiplicities.
If the eigenvalues are continuous the
 sum is replaced by an integral.
In this regularization, the gap equation \eqn{gap2} on $R^3$ is,
\beq \lim_{s\rightarrow 1}[{1 \over {\lambda (s)}}
=b^2+ \int {d^3k \over (2 \pi)^3 (k^2+m^2)^s}
=b^2+{1\over 2 {\pi}^2}
\int\limits_0^\infty dt
{t^{s-1} \over \Gamma(s)} {\int k^2 dk e^{-(k^2+m^2)t}}] . \eeq
The regularized coupling $\Lambda/\lambda(\Lambda)$
has been replaced by
$1/\lambda(s)$ in the zeta function regularization.
Also, we have used the Mellin transform to analytically continue the
zeta function. (Note that $\zeta_g(s)$ has no pole at $s=1$.)
It is easy to simplify the above gap equation to,
$$\lim\limits_{s\rightarrow 1} {1 \over {\lambda (s)}}=b^2+{m \over 4 \pi}.$$
It is well known$^{2}$ in the case of $M=R^3$ that at the non-trivial
(UV stable) fixed point,
$m=0$ and $b=0$. $m$ and $b$ are physical quantities and
are therefore regularization independent.
This implies that, in the zeta function regularization,
$\lim\limits_{s\rightarrow 1} {1/{\lambda_c(s)}}=0$.
We will be using this value of the critical coupling henceforth.

To the leading order in ${1/N}$, the
critical value of free energy density in this regularization scheme is
$$W_c[g,m^2]={N \over 2} {\rm Tr}\,\log\,(-\square_g+m^2)
=-{N \over 2} \zeta_g^{\prime}(0,m^2).$$
($(-\square_g+m^2)$ is a positive operator in our case.)
This is a
conformally invariant quantity in 3--dimensions.
We give an outline of the proof of this in odd dimensions$^{4}$,
in our paper Ref.~1.
\vglue 0.6cm
\line{\elevenbf 4. Study of the $O(N)$ sigma model on manifolds
of the type $\Sigma \times R$ \hfil}
\vglue 0.3cm
In the next three subsections we will be studying the $O(N)$ non-linear
sigma model on
 a manifold $M=\Sigma \times R$ of constant curvature.
This is of particular interest to us as
it describes a quantum phase transition at
 zero temperature. We first discuss the general form of
the gap equations on such manifolds before going on to study specific cases of
$\Sigma$.
The Green's function we need,
$G_{s}(x, x; m^2,g)=\langle x|(-\square_g+ m^2)^{-s}|x \rangle$,
can be written in terms of the geometry of the
two-dimensional manifold $\Sigma$:
\beq G_{s}(x, x; m^2,g)
=\zeta_{\Sigma \times R}(s, x, m^2)
={1 \over {\sqrt {4 \pi}}} {\Gamma(s-{1 \over 2}) \over \Gamma(s)}
\zeta_{\Sigma}(s-{1 \over 2},x,m^2)\eeq
where $
\zeta_{\Sigma}(s,x,m^2)=\langle x|[-\nabla_{\Sigma}^2+{\xi R}+m^2]^{-s}|x
\rangle.$
At the critical point, in the limit $s \rightarrow 1$, the gap equations
are given by,
\beq (-\square_g+m^2) b=0\eeq\label{zeta1}
\beq b^2=-\lim_{s \rightarrow 1} \half\zeta_\Sigma(s-\half,x,m^2).
\eeq\label{zeta2}
The zeta function, $\zeta_\Sigma(s)$,
is analytic at $s=\half$ so that the above
equation is finite as ${s \rightarrow 1}$. To evaluate
$\zeta_{\Sigma}(s-{1 \over 2},x,m^2)$, we have to find the spectrum of the
operator, $-\nabla_{\Sigma}^2+\xi R+m^2$, on the space $\Sigma$.
\vglue 0.6cm
\line{\elevenit 4.1. $S_{\rho}^1 \times S_{\rho}^1 \times R$:
Example of a space of zero curvature \hfil}
\vglue 0.3cm
This is a space of zero curvature.
The Ricci scalar $R=0$. $\rho$ denotes the radii of the two
circles.
The spectrum of
the conformal laplacian is,
{}~${\rm Sp}(-\square_g)={4 \pi^2 \over \rho^2}(p^2+q^2)+k^2$ where,
$p,q=0, \pm 1, \pm2, \cdots$ and $k$ takes values on the real line.
At the critical point, on $S^1 \times S^1 \times R$, the gap equations
\eqn{zeta1} and \eqn{zeta2} take the form,
\beq m^2 b=0 \eeq \label{gaps1s1r1}
\beq \lim_{s \rightarrow 1} {1 \over {\sqrt {4 \pi}} \rho^2}
\int\limits_0^\infty {dt \over \Gamma(s)} \quad
t^{s-{3 \over 2}} e^{- m^2 t}
\Bigl[\sum_{p=-\infty}^\infty e^{-{4 \pi^2 p^2 \over \rho^2}t}\Bigr]^2
=-b^2.\eeq\label{gaps1s1r2}
We have used the Mellin transform of the zeta function,
$\zeta_{S_\rho^1 \times S_\rho^1}(s-{1 \over 2})$, in equation
\eqn{gaps1s1r2}.
We use the standard
Poisson sum formula, $$\sum\limits_n e^{-({4 \pi^2 n^2\over{\beta}^2})t}
= {{\beta}\over {(4 \pi t)^{1 \over 2}}}\sum_n e^{-n^2 {\beta}^2\over {4 t}}
 ,$$ to extract the divergence in the sum from
the small $t$
region and perform the resulting
integrals to reduce \eqn{gaps1s1r2} to,
\beq -{1 \over 4}-{1 \over m \rho} \,\log\,(1-e^{-m \rho})
+ \sum_{p,q=1}^\infty {{e^{-{\sqrt (p^2+q^2)} m \rho}} \over m \rho
{\sqrt {p^2+q^2}}}=-{ \pi b^2 \over m}. \eeq\label{doublesum}
\n The double sum we are left with
is not an obvious one. Nevertheless, we can see that $m\not=0$
without actually solving
the equation. To see this, we
put in an ansatz $m \rightarrow 0$ and we will show that this is
inconsistent with the gap equation.
In the limit $m \rightarrow 0$, the double sum can be approximated
by a double integral which can be easily performed and the gap equation
simplifies to
a transcendental
equation for $m \rho$,
\beq {(m \rho)^2 \over 4} +(m \rho) \,\log \,{m \rho}
-{\pi \over 2}={ \pi b^2 m \rho^2}.\eeq
It is apparent from the above equation that $m=0$ cannot be one
of its solutions. This implies that $b$ has to be zero at the
critical point in order to satisfy the gap equation \eqn{gaps1s1r1}.
The critical value of $m$ as a function of $\rho$ is given by
the solution to \eqn{doublesum} with $b=0$.
With this value of $m$, the free energy density can
be computed.
\vglue 0.6cm
\line{\elevenit 4.2. $S_\rho^2 \times R$: Example of
a space of constant positive curvature \hfil}
\vglue 0.3cm
The fact that $S^2$ has finite volume might lead one to expect $m$
to be non-zero at criticality. But we will see that
this is not the case; the manifold $S^2 \times R$ is conformal to $R^3-\{0\}$
and it turns out that in fact $m=0$.

%\n On $S^2 \times R$,  $\xi R={1 \over 4 \rho^2}$ and the
On $S^2 \times R$,  $\xi R={1/ 4 \rho^2}$. The
conformal laplacian has the spectrum,
{}~${\rm Sp}~(-\square_{S^2 \times R})=[{(l+{1 \over 2})^2 \over \rho^2}+k^2]$
with degeneracy $(2l+1)$, where $l=0,1,2,\cdots$
and $k \in R$.
Notice that the conformal laplacian on $S^2 \times R$ has no zero modes.
 At the critical point, the gap equations are,
\beq({1 \over 4 \rho^2}+m^2) b =0\eeq\label{gaps2r1}
\beq -{\lim_{s\rightarrow 1}}
{1 \over {\sqrt {4 \pi}}}
\int\limits_0^\infty dt {t^{s-{3 \over 2}}\over \Gamma(s)} e^{- m^2 t}
\sum_{l={1\over 2}}^\infty 2 l e^{-{l^2 \over \rho^2}t}=b^2.\eeq
\label{gaps2r2}
Again in equation \eqn{gaps2r2} we have used the Mellin transform of
$\zeta_{S_\rho^2}(s-{1 \over 2})$.
\n From equation \eqn{gaps2r1} we see that $b=0$ since
$m^2+ {1 \over 4 \rho^2}$
cannot be equal to zero, both $m$ and $\rho$ being positive.
The sum in \eqn{gaps2r2} is divergent in the small $t$ region and we
therefore have to separate out the divergent piece in the sum before we can
interchange the sum and the integral over $t$.
To do this we use an analog of the Poisson sum formula$^{1}$ for the
case of the sum over half-integers $l$, and
the gap equation simplifies to,
\beq{\rho \over  {\sqrt \pi}}
{\rm P}\int\limits_{-\infty}^\infty dx \quad
({x} {\rm cosec}{x}-1) \int\limits_0^\infty dt \quad
t^{-2} e^{-{x^2 \rho^2 \over t}-m^2 t}
+m \Gamma(-{1 \over 2})=0\eeq\label{solvem}

\n where, by ${\rm P}\int\limits_{-\infty}^\infty dx f(x)$
we mean the principal value of the integral;
$({x} {\rm cosec}{x}-1)$
has simple poles at all non-zero
integral multiples of $\pi$.

It is easy to check$^{1}$ that the ansatz
$m=0$ in the l.h.s.\ of the above gap equation is indeed
a consistent solution.
Hence, at the critical point, both $m$ and $b$ are zero on $S^2 \times R$.
It is important to note though that even though $m=0$ at criticality,
the correlation length of
this system is not infinite. The correlation
function given by $\langle\phi_i(x,u)\phi_j(y,0)\rangle$, which is a function
of
$x,y\in S^2$ and $u\in R$, decays like an
exponential in  $u$ as $u\to\infty$ ($\sim e^{-{\sqrt {\xi R}} u}$).
(In the $x,y$ directions the space has finite size.)
This is because the operator $-\nabla_{S^2}^2+\xi R+m^2$ has no zero modes.

The regularized free energy density on  $S^2 \times R$,
at the critical point,  is
given by,
$ {W_c(\rho) \over N}=
-{1 \over 2}\zeta_{S^2 \times R}^{\prime}(0,m^2)$ and it
can be shown$^{1}$ to be zero
at the critical point ($m=0$).
This just means that the regularized free energy
density on $S^2 \times R$ is the same as that on $R^3$ at the critical point,
which is what we should expect from general considerations of conformal
equivalence of the spaces $S^2 \times R$ and $R^3-\{0\}$.
\vglue 0.6cm
\line{\elevenit 4.3. $H^2 \times R$: Example of
a space of constant negative curvature \hfil}
\vglue 0.3cm

$H^2$ is a two--dimensional hyperboloid. On the space $H^2 \times R$,
$\xi R=-{1/4 \rho^2}.$
Let the co-ordinate on $R$ be denoted by $u$ and the hyperboloid be
parametrized as follows:
$H^2=\{z=(x,y):x \in R, 0<y<\infty\}$
with line element,
$ds^2={\rho^2 \over y^2} (dx^2+dy^2)$ and laplacian
{}~$\nabla_{H^2}^2={y^2 \over \rho^2} (\partial_x^2+\partial_y^2).$
The spectrum of the operator
$-\nabla_{H^2}^2+\xi R+m^2$ on the hyperbolic space
is continuous and therefore we need the density of states
$\rho(\lambda)$ to be able to evaluate
$\zeta_{H^2}(s-{1 \over 2})=[{\rm Tr}(-\nabla_{H^2}^2-{1 \over 4 \rho^2}+m^2]
^{-(s-{1 \over 2})}$. $\rho(\lambda)$ is found$^{5,1}$ to be,
\beq
\rho(\lambda)={1 \over 8 \pi}\Theta(-{1 \over 4}+\lambda)
\tanh{\pi {\sqrt{-{1 \over 4}+\lambda}}}.\eeq
(For simplicity, we assume that $\rho=1$).
The gap equations at the critical point are,
\beq (-{1 \over 4 \rho^2}+m^2) b=0\eeq\label{gaph2r1}
\beq -{\lim_{s \rightarrow 1}}
{1 \over 2}
\int d\lambda \quad \rho(\lambda)
[\lambda-{1 \over 4 \rho^2}+m^2]^{-(s-{1 \over 2})}=b^2. \eeq \label{gaph2r2}
On using the Mellin transform of the zeta function and
after some simplification, the gap equation \eqn{gaph2r2} reduces to,
\beq
m +\int\limits_0^\infty dk \quad
{k \over {\sqrt{k^2+m^2}}} (1-\tanh{\pi k})=b^2.\eeq\label{solveb}

\n Observe that each term in
the l.h.s.\ is manifestly positive. $b=0$ cannot therefore be
a solution if $m\not=0$. If $m=0$, it is easy to check that the
l.h.s.\ is non-zero (it is equal to $(\ln2)/ \pi$)
and therefore $b \neq 0$.
In order to satisfy the gap equation \eqn{gaph2r1}, we
require $m={1/ 2 \rho}$ (since we set $\rho=1$, $m={1 \over 2}$)
at the critical point. The non-zero value of the order parameter $b$
gives the non-zero spontaneous magnetization in the system ---
negative curvature has induced the symmetry to be
spontaneously broken at
the critical point.
 At criticality, from
\eqn{solveb}, we find,
\beq b^2\rho={{\sqrt \pi}\over 2} [1 +
\sum_{r=0}^\infty {{1 \over 2} \choose r}
{2\over \pi^{2r+2}} (2r+1)! (1-2^{-2r-1}) \zeta(2r+2)].\eeq

Notice, $-\nabla_{H^2}^2+\xi R+m^2$ has a zero mode
for the critical value $m={1/2\rho}$.
This means that the correlation
functions fall off with distance like a power law,
not an exponential. Thus even though $m$
is non--zero, the correlation length is in fact
infinite on $H^2 \times R$.

%\n The
The
critical value of the free energy density on $H^2 \times R$ works out to be,
\beq {W_c(\rho) \over N}
= -{1 \over 2} \zeta_{H^2 \times R}^{\prime}(0, \rho)
= -{1 \over 24 \rho^3}
-{1 \over 4 \rho^3} \sum_{r=0}^\infty {{1 \over 2} \choose r}
{1 \over \pi^{2r+2}} (2r+1)! (1-2^{-2r-1}) \zeta(2r+2).\eeq

\vglue 0.3cm
\n {\elevenbf References}
\vglue 0.3cm
\item {1.} S. Guruswamy, S.G. Rajeev and P. Vitale,
`$O(N)$ Sigma Model as a Three
Dimensional Conformal Field Theory', {\elevenit Preprint UR-1357,
hep-th/9406010} (and references therein).
\item {2.} A.M. Polyakov, {\elevenit Gauge fields and strings}, (Harwood
Academic Publishers, 1987).
\item {3.} S. Sachdev, {\elevenit Phys. Lett.} {\elevenbf B309} (1993) 285;
B. Rosenstein, B.J. Warr and S.H. Park, {\elevenit Nucl.
Phys.} {\elevenbf B116} (1990) 435.
\item {4.} T. Parker and S. Rosenberg, {\elevenit J. Differential Geometry}
 {\elevenbf 25} (1987) 199.
\item {5.} S. Lang, {\elevenit $SL_2(R)$}, (Springer-Verlag, 1985).

\end